\documentclass
[superscriptaddress,secnumarabic,amssymb,amsmath,nobibnotes,aps,prd,showkeys,showpacs,nofootinbib,onecolumn,12pt]{revtex4}%
\usepackage{graphicx}
\usepackage{epsf}
\usepackage{bm}
\usepackage{amsmath}
\usepackage{amsfonts}
\usepackage{amssymb}%
\providecommand{\U}[1]{\protect\rule{.1in}{.1in}}

\newcommand{\be}{\begin{equation}}
\newcommand{\ee}{\end{equation}}

\newcommand{\mincir}{\raise
-3.truept\hbox{\rlap{\hbox{$\sim$}}\raise4.truept\hbox{$<$}\ }}
\newcommand{\magcir}{\raise
-3.truept\hbox{\rlap{\hbox{$\sim$}}\raise4.truept\hbox{$>$}\ }}

\begin{document}
\title{Szekeres Universes with GUP corrections}
\author{Andronikos Paliathanasis}
\email{anpaliat@phys.uoa.gr}
\affiliation{Institute of Systems Science \& Department of Mathematics, Faculty of Applied
Sciences, Durban University of Technology, Durban 4000, South Africa}
\affiliation{Centre for Space Research, North-West University, Potchefstroom 2520, South Africa}
\affiliation{Departamento de Matem\'{a}ticas, Universidad Cat\'{o}lica del Norte, Avda.
Angamos 0610, Casilla 1280 Antofagasta, Chile}

\begin{abstract}
We demonstrate that introducing a deformed algebra with a minimum length
modifies the field equations for an inhomogeneous spacetime, resulting in the
emergence of acceleration. Specifically, we examine the analytic effects of
the Generalized Uncertainty Principle on the classical field equations of the
Szekeres system. Our findings show that the deformed algebra leads to a
modified Szekeres system capable of describing cosmic acceleration. Moreover,
the spatial curvature of the spacetime is influenced by the presence of the
minimum length.

\end{abstract}
\date{\today}
\maketitle

\section{Introduction}

Scalar field inflation is the main mechanism used to explain the homogeneity
and isotropy of the present universe. The scalar field dominates in driving
the dynamics and explaining the expansion era \cite{Aref1,guth}. Nevertheless,
scalar field inflationary models are primarily defined on homogeneous
spacetimes or background spaces with small inhomogeneities \cite{st1,st2}. In
\cite{w1}, it was found that the presence of a positive cosmological constant
in Bianchi cosmologies leads to expanding Bianchi spacetimes evolving toward
the de Sitter universe. This was the first result supporting the cosmic
\textquotedblleft no-hair\textquotedblright\ conjecture \cite{nh1,nh2}. The
conjecture posits that all expanding universes with a positive cosmological
constant asymptotically admit the de Sitter universe as a solution.

The necessity of de Sitter expansion is that it provides a rapid expansion of
the universes size, causing the universe to effectively lose memory of its
initial conditions. This means that de Sitter expansion resolves the
\textquotedblleft flatness,\textquotedblright\ \textquotedblleft
horizon,\textquotedblright\ and monopole problems \cite{f1,f2}. The
inhomogeneous silent spacetimes, known as Szekeres universes \cite{szek0},
with a cosmological constant, were investigated in \cite{ba1}, where it was
found that the cosmological constant supports the cosmic \textquotedblleft
no-hair\textquotedblright\ conjecture. The validity of the conjecture is also
examined in \cite{sim1}. Szekeres spacetimes are exact solutions of General
Relativity with an inhomogeneous pressureless fluid source, where the metric
tensor typically depends on two scale factors. These two scale factors are
inhomogeneous, meaning that, in general, Szekeres spacetimes do not possess
any symmetries \cite{mm1}. For this physical scenario, Einstein's field
equations provide two families of solutions: one in which the two scale
factors evolve identically over time, governed by the field equations of FLRW
geometry, and another where the two scale factors evolve according to the
anisotropic Kantowski-Sachs geometry. LTB spacetimes \cite{ltb2,ltb3} and
T-models \cite{novikov,tm} are examples of exact solutions belonging to the
Szekeres family of spacetimes.

Szekeres spacetimes belong to the family of \textquotedblleft
Silent\textquotedblright\ universes \cite{mi1}. Specifically, the fluid source
lacks a pressure component, and the magnetic part of the Weyl tensor is zero.
As a result, there is no information transfer between different cosmic lines
via sound or gravitational waves. Numerous applications in the literature
discuss the significance of Szekeres spacetimes
\cite{bon,Gol,grpan1,grpan2,zeldpan}. A generalization of the Szekeres model
was introduced in \cite{szafron}, where a pressure term was incorporated into
the fluid source. Further generalizations of the Szekeres geometries can be
found in \cite{kraj}.

Various approaches to quantum gravity, such as string theory \cite{ml1},
double special relativity \cite{ml2}, noncommutative geometry \cite{ml3}, and
loop quantum gravity \cite{ml4}, predict the existence of a minimum length
scale on the order of the Planck length ($l_{PL}$). This prediction leads to a
modification of Heisenberg's Uncertainty Principle, known as the Generalized
Uncertainty Principle (GUP), in the context of quantum gravity \cite{Maggiore}%
. For an extended review of GUP, we refer the reader to \cite{sb1}. In the
case of a homogeneous and isotropic universe, the effects of GUP have been
used to modify the field equations \cite{sb2}, while the relation between
inflationary physics and GUP can be found in \cite{sb3,sb4}. There are
numerous applications of GUP in gravitational physics; see, for instance,
\cite{gg1,gg2,gg3,gg4,gg5,gg6,gg7,gg8,gg9,gg10,gg11} and references therein.

The time evolution of the physical variables for the Szekeres model, as
described by the gravitational field equations, forms a system of four
first-order ordinary differential equations known as the Szekeres system.
Meanwhile, the spatial evolution is governed by a set of constraint equations.
The structure and integrability properties of the Szekeres system have been
the subject of various studies. In \cite{anszek}, it was determined that the
Szekeres system follows from a variational principle. Specifically, a
minisuperspace description exists, and a canonical Hamiltonian function can be
constructed. This property has been used previously to define the
Wheeler-DeWitt equation of quantum cosmology and to study the semiclassical
limit of Szekeres spacetimes.

In this study, we are interested in investigating the effects of a minimum
length on the field equations of inhomogeneous Szekeres spacetimes. We examine
the evolution of the physical variables under the application of the new
deformed algebra. The structure of the paper is as follows:

In Section \ref{sec2}, we introduce the Szekeres models and discuss the
Hamiltonian formalism for the Szekeres system. The basic elements of GUP are
provided in Section \ref{sec3}. The effects of the minimum length on the
Szekeres system are discussed in Section \ref{sec4}, where we rewrite the
field equations and determine the new physical properties of the asymptotic
solutions of the new dynamical system. Finally, in Section \ref{sec5}, we
present our conclusions.

\section{Szekeres Cosmologies}

\label{sec2}

We introduce the inhomogeneous spacetime \cite{szek0}
\begin{equation}
ds^{2}=-dt^{2}+e^{2a\left(  t,x,y,z\right)  }dx^{2}+e^{2b\left(
t,x,y,z\right)  }\left(  dx^{2}+dy^{2}+dz^{2}\right)  , \label{ss.00}%
\end{equation}
where the scale factors $a\left(  t,x,y,z\right)  ,~b\left(  t,x,y,z\right)  $
satisfy Einstein's field equations%
\begin{equation}
R_{\mu\nu}-\frac{1}{2}Rg_{\mu\nu}=T_{\mu\nu}.
\end{equation}
The tensor $T_{\mu\nu}$ describes the pressureless and inhomogeneous matter
term of the universe, that is%
\[
T_{\mu\nu}=\rho\left(  t,x,y,z\right)  u_{\mu}u_{\nu},
\]
in $u^{\mu}=\delta_{t}^{\mu}$ is the comoving observer.

The time evolution of the scale factors $a\left(  t,x,y,z\right)  ,~b\left(
t,x,y,z\right)  $ and of the matter density $\rho\left(  t,x,y,z\right)  $ is
given by the known Szekeres system \cite{kraj}%
\begin{align}
\dot{\rho}+\theta\rho &  =0,~\label{ss.01}\\
\dot{\theta}+\frac{\theta^{2}}{3}+6\sigma^{2}+\frac{1}{2}\rho &
=0,\label{ss.02}\\
\dot{\sigma}-\sigma^{2}+\frac{2}{3}\theta\sigma+\mathcal{E}  &
=0,\label{ss.03}\\
\mathcal{\dot{E}}+3\mathcal{E}\sigma+\theta\mathcal{E}+\frac{1}{2}\rho\sigma
&  =0, \label{ss.04}%
\end{align}
with the constraint
\begin{equation}
\frac{\theta^{2}}{3}-3\sigma^{2}+\frac{^{\left(  3\right)  }R}{2}=\rho,
\label{ss.05}%
\end{equation}
where $\dot{}$ remarks the directional derivative along $u^{\mu}$, i.e.
$\dot{}=u^{\mu}\nabla_{\mu}$.

The parameter $^{\left(  3\right)  }R$ denotes the curvature of the
three-dimensional hypersurfaces, while $\theta$ represents the expansion rate,
and $\sigma$ and $\mathcal{E}$ refer to the shear and the electric component
of the Weyl tensor, respectively. These inhomogeneous quantities are related
to the functions $\alpha$, $b$, and their derivatives. For more details, we
refer the reader to \cite{kraj}.

The space-evolution of the physical parameters is given by the constraint
equations
\begin{align}
h_{\mu}^{\nu}\sigma_{\nu;\alpha}^{\alpha}  &  =\frac{2}{3}h_{\mu}^{\nu}%
\theta_{;\nu},~\\
h_{\mu}^{\nu}E_{\nu;\alpha}^{\alpha}  &  =\frac{1}{3}h_{\mu}^{\nu}\rho_{;\nu}%
\end{align}
in which $h_{\mu\nu}$ is the decomposable tensor defined by the expression
\begin{equation}
h_{\mu\nu}=g_{\mu\nu}-\frac{1}{u_{\lambda}u^{\lambda}}u_{\mu}u_{\nu}.
\end{equation}

The deceleration parameter is defined as $q=-1-3\frac{\dot{\theta}}{\theta
^{2}}$, that is,%
\begin{equation}
q=18\left(  \frac{\sigma}{\theta}\right)  ^{2}+\frac{3}{2}\frac{\rho}%
{\theta^{2}}. \label{ss.05a}%
\end{equation}

\subsection{Hamiltonian formalism}

The four first-order differential equations that constitute the Szekeres
system may describe a Hamiltonian system with $2\times2$ degrees of freedom.
Following \cite{anszek}, we introduce a new set of variables with the
following transformation rule%

\begin{equation}
\rho=\frac{6}{u^{2}\left(  u-v\right)  },~\mathcal{E}=-\frac{v}{u^{3}\left(
u-v\right)  }%
\end{equation}%
\begin{equation}
~\sigma=\frac{v\dot{u}-u\dot{v}}{3u\left(  v-u\right)  },~\theta=\frac{\left(
3u-2v\right)  \dot{u}-u\dot{v}}{u\left(  u-v\right)  },
\end{equation}
the, in terms of the new coordinates the Szekeres system (\ref{ss.01}),
(\ref{ss.02}), (\ref{ss.03}) and (\ref{ss.04}) is written as follows%
\begin{equation}
\dot{u}=p_{v}~,~\dot{v}=p_{u},
\end{equation}%
\begin{equation}
\dot{p}_{u}=2\frac{v}{u^{3}}~,~\dot{p}_{v}=-\frac{1}{u^{2}}.
\end{equation}

We observe that the latter system are the Hamilton's equations for the
\begin{equation}
h=p_{u}p_{v}+\frac{v}{u^{2}}, \label{ss.06}%
\end{equation}
and%
\begin{equation}
\dot{u}=\frac{\partial H}{\partial p_{u}},~\dot{v}=\frac{\partial H}{\partial
p_{v}},~\dot{p}_{u}=-\frac{\partial H}{\partial u},~\dot{p}_{v}=-\frac
{\partial H}{\partial v}.
\end{equation}

In terms of the original variables, the Hamiltonian function (\ref{ss.06})
reads%
\begin{equation}
h=\frac{1}{81}\rho^{\frac{2}{3}}e^{Z}\left(  6p_{\rho}-p_{Z}\right)  \left(
3p_{\rho}+p_{Z}\right)  +\frac{1}{2}\rho^{\frac{1}{2}}e^{-4Z}\left(
3-9e^{3Z}\right)  , \label{ss.07}%
\end{equation}
where$\mathcal{~E=}\frac{1}{6}\left(  e^{-3Z}-1\right)  \rho$, and
\begin{equation}
\theta=-\dot{\rho}~,~\sigma=\dot{Z}. \label{ss.08}%
\end{equation}

We calculate that $\theta$ and $\sigma$ are expressed in terms of the momentum
$p_{\rho}$,~$p_{Z}$ as follows
\begin{align}
\theta &  =\frac{1}{27}e^{Z}\rho^{\frac{2}{3}}\left(  12p_{\rho}+p_{Z}\right)
,\label{ss.09}\\
\sigma &  =\frac{1}{81}e^{Z}\rho^{\frac{2}{3}}\left(  3p_{\rho}-2p_{Z}\right)
, \label{ss.10}%
\end{align}
or equivalently%
\begin{align}
p_{\rho}  &  =e^{-Z}\rho^{-\frac{2}{3}}\left(  2\theta-3\sigma\right)
,\label{ss.11}\\
p_{Z}  &  =-3e^{-Z}\rho^{-\frac{2}{3}}\left(  \theta+12\sigma\right)  .
\label{ss.12}%
\end{align}

Furthermore, the rest of Hamilton's equations are%
\begin{equation}
\dot{p}_{\rho}=-\left(  \frac{1}{6}e^{-4Z}\rho^{\frac{1}{3}}\left(
3-9e^{3Z}\right)  +\frac{2}{243}e^{Z}\rho^{\frac{2}{3}}\left(  6p_{\rho}%
-p_{Z}\right)  \left(  3p_{\rho}+p_{Z}\right)  \right)  , \label{ss.14}%
\end{equation}%
\begin{equation}
\dot{p}_{z}=-\left(  \frac{1}{81}\rho^{\frac{2}{3}}e^{Z}\left(  6p_{\rho
}-p_{Z}\right)  \left(  3p_{\rho}+p_{Z}\right)  +\frac{3}{2}e^{-4Z}\rho
^{\frac{1}{3}}\left(  -4+3e^{3Z}\right)  \right)  . \label{ss.15}%
\end{equation}

\section{Generalized Uncertainty Principle}

\label{sec3}

We introduce a minimum measurable length, which modifies the structural form
of the uncertainty principle. In the so-called quadratic GUP, Heisenberg's
uncertainty principle becomes
\begin{equation}
\Delta X_{i}\Delta P_{j}\geqslant\frac{\hbar}{2}[\delta_{ij}(1+\beta
P^{2})+2\beta P_{i}P_{j}], \label{gp.01}%
\end{equation}
where $\beta= \frac{\beta_{0}}{M_{Pl}^{2}c^{2}} = \frac{\beta_{0} \ell
_{Pl}^{2}}{2\hbar^{2}}$, $\beta_{0}$ is the deformation parameter, $M_{Pl}$ is
the Planck mass, $\ell_{Pl}$ is the Planck length, and $M_{Pl}c^{2}$ is the
Planck energy.

From the uncertainty principle (\ref{gp.01}) follows the deformed Heisenberg
algebra \cite{Vagenas,Moayedi},
\begin{equation}
\lbrack X_{i},P_{j}]=i\hbar\lbrack\delta_{ij}(1-\beta P^{2})-2\beta P_{i}%
P_{j}]. \label{md1}%
\end{equation}
The coordinate representation of the modified momentum operator is
$P_{i}=p_{i}(1-\beta p^{2})$ \cite{Moayedi}, while keeping $X_{i}=x_{i}$
undeformed. Thus the pair of variables $\left(  x_{i},~p_{i}\right)  ~$define
the canonical representation satisfying $[x_{i},p_{j}]=i\hbar\delta_{ij}%
$\thinspace$,$ and $p^{2}=\gamma^{\mu\nu}p_{\mu}p_{\nu}$.

The deformed Heisenberg algebra leads to the modification of the Hamilton's
equations as follows \cite{hm1,hm2,hm3}%
\begin{equation}
\dot{x}^{i}=\left(  1-\beta p^{2}\right)  \frac{\partial H}{\partial p_{i}%
}~,~\dot{p}^{i}=\left(  1-\beta p^{2}\right)  \frac{\partial H}{\partial
x_{i}}\text{.}%
\end{equation}

The quadratic GUP is not the only modification of the uncertainty principle
that has been proposed in the literature \cite{hm4,hm5,hm6,hm7,hm8}. Various
modifications exist and can be summarized by the following general family of
uncertainty principles:%

\begin{equation}
\Delta X_{i}\Delta P_{j}\geqslant\frac{\hbar}{2}[\delta_{ij}(1+\beta f\left(
P\right)  )].
\end{equation}
Consequently, the modified Hamilton's equations read%
\begin{equation}
\dot{x}^{i}=\left(  1-\beta f\left(  p\right)  \right)  \frac{\partial
H}{\partial p_{i}}~,~\dot{p}^{i}=\left(  1-\beta f\left(  p\right)  \right)
\frac{\partial H}{\partial x_{i}}. \label{md3}%
\end{equation}

In the following Section we employ the deformed algebra to find the modified
Szekeres system in the presence of the minimum length.

\section{Szekeres system with deformed algebra}

\label{sec4}

We consider the deformed algebra given by (\ref{md3}). Therefore, from
Hamilton's equations for the function (\ref{ss.07}), we determine the modified
Szekeres system
\begin{align}
\left(  \dot{\rho}+\rho\theta\right)  +\beta f\left(  p\right)  \rho\theta &
=0,\label{ms.01}\\
\left(  \dot{Z}-\sigma\right)  +\beta f\left(  p\right)  \sigma &
=0,\label{ms.02}\\
\left(  \dot{\theta}+\frac{\theta^{2}}{3}+6\sigma^{2}+\frac{1}{2}\rho\right)
+\beta f\left(  p\right)  \left(  \frac{\rho}{2}-\frac{\theta^{2}}{3}%
-6\sigma^{2}\right)   &  =0,\label{ms.03}\\
\left(  \dot{\sigma}-\sigma^{2}+\frac{2}{3}\theta\sigma+\frac{\rho}{6}\left(
e^{-3Z}-1\right)  \right)  +\beta f\left(  p\right)  \frac{\rho}{3}\left(
e^{-3Z}-1\right)   &  =0. \label{ms.04}%
\end{align}
We remark that in the limit where $\beta f\left(  p\right)  \rightarrow0$, we
recover the underformed Szekeres system. The new dynamical terms introduced by
the GUP are small perturbations which however can affect the physical
properties of the general dynamics. In terms of the electric component
$\mathcal{E}$, equation (\ref{ms.04}) is expressed as%
\begin{equation}
\left(  \dot{\sigma}-\sigma^{2}+\frac{2}{3}\theta\sigma+\mathcal{E}\right)
+2\beta f\left(  p\right)  \mathcal{E}=0. \label{ms.05}%
\end{equation}

Indeed, the deceleration parameter (\ref{ss.05a}) is modified as
\begin{equation}
\tilde{q}=-\beta f\left(  p\right)  +18\left(  \frac{\sigma}{\theta}\right)
^{2}\left(  1-\beta f\left(  p\right)  \right)  +\frac{3}{2}\frac{\rho}%
{\theta^{2}}\left(  1+\beta f\left(  p\right)  \right)  . \label{ms.06}%
\end{equation}
Thus, in the isotropic and vacuum case, where $\sigma=0$, $\rho=0$, and
$\mathcal{E}=0$, where the classical Szekeres system describe an inhomogeneous
Milne (-like) geometry with $q=0$, after the application of the deformed
algebra we calculate $\tilde{q}=-\beta f\left(  p\right)  $, which is negative
when $\beta f\left(  p\right)  >0$. Consequently, the GUP can describe an
inflationary mechanism for the inhomogeneous cosmological model. Although
$\beta$ is small, when in the Milne solution $f\left(  p\right)  $ is a
singular function such as $f\left(  p\right)  \simeq\beta^{-1}$, then the
deformed algebra rides the rapid expansion of the universe, and when $\beta
f\left(  p\right)  \rightarrow-1$, the de Sitter universe is recovered.
Furthermore, either in anisotropic and vacuum solutions, i.e. $\rho
\rightarrow0$, $\sigma\neq0$ if $f\left(  p\right)  $ has a singular behaviour
such that $\beta f\left(  p\right)  \simeq1$, then the corresponding universe
describes acceleration. We conclude that the introduction of the deformed
Heisenberg algebra modifies the Szekeres system, resulting in the emergence of
inflationary solutions. As we shall see in the following lines, even a simple
modification of the Heisenberg algebra by a constant parameter, i.e.,
$f\left(  p\right)  = f_{0}$, leads to an attractor for the Szekeres system
that describes acceleration.

Consider now that $f\left(  p\right)  $ is constant, without loss of
generality we assume that $f\left(  p\right)  =1$. We introduce the new
dimensionless variables%
\begin{equation}
\Omega_{m}=3\frac{\rho}{\theta^{2}},~\Sigma=\frac{\sigma}{\theta}%
,~A=\frac{\mathcal{E}}{\theta^{2}},~\Omega_{R}=\frac{3}{2}\frac{^{\left(
3\right)  }R}{\theta^{2}}. \label{ms.07}%
\end{equation}
Thus, the field equations are reduced to the following set of differential
equations%
\begin{align}
\Omega_{m}^{\prime}  &  =\frac{1}{3}\Omega_{m}\left(  36\Sigma^{2}-\Omega
_{m}-1\right)  +\frac{\beta}{3}\Omega_{m}\left(  \Omega_{m}-5-36\Sigma
^{2}\right)  ,\label{ms.08}\\
\Sigma^{\prime}  &  =\frac{1}{6}\Sigma\left(  \Omega_{m}-2+6\Sigma\left(
1+6\Sigma\right)  \right)  -A+\frac{\beta}{6}\left(  \Sigma\left(  \Omega
_{m}-2-36\Sigma^{2}\right)  -12A\right)  ,\label{ms.09}\\
A^{\prime}  &  =\frac{1}{6}\left(  2A\left(  9\Sigma\left(  4\Sigma-1\right)
+\Omega_{m}-1\right)  -\Sigma\Omega_{m}\right)  -\frac{\beta}{6}\left(
2A\left(  5+9\Sigma\left(  1+4\Sigma\right)  -\Omega_{m}\right)  +\Sigma
\Omega_{m}\right)  , \label{ms.10}%
\end{align}
where $^{\prime}$ denotes derivative with respect to the new independent
variable $dt=\theta d\tau$. Moreover, we calculation the constraint equation%
\begin{equation}
\Omega_{R}=\Omega_{m}-1+9\Sigma^{2}. \label{ms.11}%
\end{equation}
Finally, in terms of the new dimensionless variables the deceleration
parameter is expressed as follows%
\begin{equation}
\tilde{q}=18\Sigma^{2}+\frac{1}{2}\Omega_{m}-\beta\left(  1+18\Sigma^{2}%
-\frac{1}{2}\Omega_{m}\right)  . \label{ms.12}%
\end{equation}

In order to understand the evolution of the physical parameters in the
solution space, we determine the stationary points for the dynamical system
(\ref{ms.08}), (\ref{ms.09}), and (\ref{ms.10}). At each stationary point,
there corresponds a specific asymptotic solution. The analysis of dynamical
systems has been widely applied in gravitational physics. In the case of the
silent universe and for the Szekeres spacetimes, phase-space analysis is
presented in \cite{mi1}. It has been found that the Szekeres system
(\ref{ss.01}), (\ref{ss.02}), (\ref{ss.03}), and (\ref{ss.04}) possesses the
inhomogeneous Milne-like universe as a unique future attractor.

In the following lines we present the stationary points $P=\left(  \Omega
_{m}\left(  P\right)  ,\Sigma\left(  P\right)  ,A\left(  P\right)  \right)  $
for the dynamical system (\ref{ms.08}), (\ref{ms.09}) and (\ref{ms.10}), and
we discuss\ their stability properties, and the physical character of the
corresponding solutions. In the following we assume that $\beta\neq0$ and
$O\left(  \beta^{2}\right)  \rightarrow0$.

Specifically we calculate:
\[
P_{1}=\left(  0,0,0\right)  ,
\]
with $\Omega_{R}\left(  P_{1}\right)  =-1$, describes a FLRW-like geometry
with negative spatially curvature. The deceleration parameter is $\tilde
{q}\left(  P_{1}\right)  =-\beta,$ from where it follows that $\tilde{q}<0$
when $\beta>0$. In the limit where $\beta$ is zero the Milne-like universe is
recovered. The eigenvalues of the linearized system around the stationary
point $P_{1}$ are $-\frac{\left(  1+5\beta\right)  }{3}$,~$-\frac{\left(
1+5\beta\right)  }{3}$,~$-\frac{\left(  1+\beta\right)  }{3}$; thus $P_{1}$ is
always an attractor, recall that $\beta^{2}\rightarrow0$.%
\[
P_{2}=\left(  1+4\beta,0,0\right)  ,
\]
where $\Omega_{R}\left(  P_{2}\right)  =4\beta$, corresponds to a FLRW-like
geometry with positive spatially curvature and nonzero matter source. The
deceleration parameter is $\tilde{q}\left(  P_{2}\right)  =\frac{1}{2}\left(
1+3\beta\right)  $, and the eigenvalues read $\frac{\left(  1+5\beta\right)
}{3}$,$~-\frac{\left(  5+11\beta\right)  }{10}$ and $\frac{\left(
15+8\beta\right)  }{5}$. Hence, $P_{2}$ is a saddle point.%

\[
P_{3}=\left(  0,\frac{1}{3}+\frac{4}{5}\beta,\frac{2}{9}+\frac{49}{45}%
\beta\right)
\]
with spatial curvature parameter $\Omega_{R}\left(  P_{3}\right)  =\frac{8}%
{5}\beta$, and deceleration parameter $\tilde{q}\left(  P_{3}\right)
=2+\frac{33}{5}\beta$. The corresponding asymptotic solution describes an
anisotropic universe with nonzero spatial curvature. The eigenvalues are
$1+\frac{17}{5}\beta$, $\frac{2}{3}+\frac{161}{45}\beta$ and $\frac{5}%
{3}+\frac{238}{45}\beta$ from where infer that point $P_{3}$ is always a
source point.%

\[
P_{4}=\left(  0,-\left(  \frac{1}{12}+\frac{3}{10}\beta\right)  ,\frac{1}%
{32}+\frac{13}{160}\beta\right)  ,
\]
and physical parameters $\Omega_{R}\left(  P_{4}\right)  =-\frac{5}{16}%
+\frac{3}{20}\beta$, $\tilde{q}\left(  P_{4}\right)  =\frac{5-9\beta}{40}$,
describes an anisotropic universe. We calculate the eigenvalues $-\frac
{5-23\beta}{4}$, $-\frac{5}{8}-\frac{1031}{840}\beta$, $\frac{1}{4}+\frac
{449}{420}\beta$, where we conclude that point $P_{4}$ is a saddle point.%

\[
P_{5}=\left(  0,-\frac{\left(  1+\beta\right)  }{3},0\right)  ,
\]
describes an anisotropic universe with $\Omega_{R}\left(  P_{5}\right)
=2\beta$, $\tilde{q}\left(  P_{5}\right)  =2+\beta$.\ $P_{5}$ is a always
source point because the corresponding eigenvalues are $1+\beta$,~$2+\frac
{5}{3}\beta$, $1-\frac{\beta}{3}$.%

\[
P_{6}=\left(  0,\frac{\left(  1+\beta\right)  }{6},0\right)  ,
\]
provides an asymptotic solution which describes an anisotropic spacetime with
$\Omega_{R}\left(  P_{6}\right)  =-\frac{3}{4}+\frac{\beta}{2}$, $\tilde
{q}\left(  P_{6}\right)  =\frac{1-\beta}{2}$, and eigenvalues $\frac{\left(
1+\beta\right)  }{2}$,~$-\frac{4}{3}\beta$, $-\frac{1}{2}-\frac{7}{3}\beta$.
Thus $P_{5}$ is always a saddle point.%
\[
P_{7}=\left(  -4\beta,\frac{1}{6}+\frac{5}{6}\beta,\frac{2\beta}{9}\right)  ,
\]
with $\Omega_{R}\left(  P_{7}\right)  =-\frac{3}{4}-\frac{3}{2}\beta$, and
$\tilde{q}\left(  P_{7}\right)  =\frac{1+3\beta}{2}$. The eigenvalues are
determined to be $-\frac{1}{2}-3\beta$, $\frac{4\beta}{3}$ and $\frac
{1+5\beta}{2}$. Thus, point $P_{7}$ is a source and describes an unstable
anisotropic universe.\ The solution is physically accepted for $\beta<0$.
\[
P_{8}=\left(  -3+4\beta,-\frac{\left(  1+\beta\right)  }{3},\frac{1}{6}%
-\frac{2}{9}\beta\right)  ,
\]
describes a non physically accepted solution.\ The stationary point is always
a source, that is, the asymptotic solution is unstable.

We remark that the dynamical system (\ref{ms.08}), (\ref{ms.09}), and
(\ref{ms.10}) admit eight stationary points, while the original Szekeres
system provides seven stationary points \cite{mi1}. The new stationary point
is $P_{7}$, which is physically accepted only for $\beta<0$. In the case where
$\beta\rightarrow0$, $P_{7}$ coincides with point $P_{6}$, and then we recover
all the stationary points for the original system \cite{mi1}. Furthermore, we
remark that the stability properties of the points remain the same.

However, the existence of the minimal length, and the introduction of the
deformed algebra, leads to different behaviour for the physical solution at
the stationary points. For instance, there is no asymptotic solution that
describes a Kasner Universe, while due to the existence of the parameter
$\beta$, there exists an asymptotic solution that can describe cosmic acceleration.

A more general consideration for the deformation function $f\left(  p\right)
$ may lead to different behaviour for the solution space. However, such
analysis is outside the scope of this study.

\section{Cosmological Constant}

The cosmological constant $\Lambda$ for the Szekeres system was introduced in
\cite{ba1}. For a negative $\Lambda$ \ the unique attractor for the Szekeres
system is the de Sitter universe.

Within the framework of GUP, the modified Szekeres equation (\ref{ms.03})
becomes%
\begin{equation}
\left(  \dot{\theta}+\frac{\theta^{2}}{3}+6\sigma^{2}+\frac{1}{2}\rho
-\Lambda\right)  +\beta f\left(  p\right)  \left(  \Lambda-\frac{\rho}%
{2}-\frac{\theta^{2}}{3}-6\sigma^{2}\right)  =0,
\end{equation}
and the constraint (\ref{ss.05}) reads%
\begin{equation}
\frac{\theta^{2}}{3}-3\sigma^{2}+\frac{^{\left(  3\right)  }R}{2}-\Lambda
=\rho.
\end{equation}
Hence, the deceleration parameter reads%
\begin{equation}
\tilde{q}_{\Lambda}=\tilde{q}-3\left(  1+\beta\right)  \frac{\Lambda}%
{\theta^{2}},
\end{equation}

In terms of the dimensionless variables (\ref{ms.07}) and $\Omega_{\Lambda
}=\frac{3\Lambda}{\theta^{2}}$ we determine the dimensionless dynamical system%
\begin{align}
\Omega_{m}^{\prime}  &  =\frac{1}{3}\Omega_{m}\left(  36\Sigma^{2}-\Omega
_{m}-1-2\Omega_{\Lambda}\right)  +\frac{\beta}{3}\Omega_{m}\left(  \Omega
_{m}-2\Omega_{\Lambda}-5-36\Sigma^{2}\right)  ,\\
\Sigma^{\prime}  &  =\frac{1}{6}\Sigma\left(  \Omega_{m}-2+6\Sigma\left(
1+6\Sigma\right)  -2\Omega_{\Lambda}\right)  -A+\frac{\beta}{6}\left(
\Sigma\left(  \Omega_{m}-2\Omega_{\Lambda}-2-36\Sigma^{2}\right)  -12A\right)
,\\
A^{\prime}  &  =\frac{1}{6}\left(  2A\left(  9\Sigma\left(  4\Sigma-1\right)
+\Omega_{m}-1-2\Omega_{\Lambda}\right)  -\Sigma\Omega_{m}\right)  -\frac
{\beta}{6}\left(  2A\left(  5+9\Sigma\left(  1+4\Sigma\right)  -\Omega
_{m}-2\Omega_{\Lambda}\right)  +\Sigma\Omega_{m}\right)  ,\\
\Omega_{\Lambda}^{\prime}  &  =\frac{1}{3}\Omega_{\Lambda}\left(
2+36\Sigma^{2}+\Omega_{m}-2\Omega_{\Lambda}\right)  +\frac{\beta}{3}%
\Omega_{\Lambda}\left(  \Omega_{m}-2\left(  1+\Omega_{\Lambda}\right)
-36\Sigma^{2}\right)  ,
\end{align}
with constraint%
\[
\Omega_{R}=\Omega_{m}+\Omega_{\Lambda}-1+9\Sigma^{2},
\]
and $\tilde{q}_{\Lambda}=\tilde{q}-\left(  1+\beta\right)  \Omega_{\Lambda}$.

The point with $\Omega_{m}=0$,~$\Sigma=0,$ $A=0$ and $\Omega_{\Lambda
}=1-2\beta$ is a stationary point for the latter dynamical system with
$\Omega_{R}=2\beta$ and $\tilde{q}_{\Lambda}=-1$. \ In the limit in which
$\beta\rightarrow0$ the asymptotic solution is reduced to the de Sitter
solution. Nevertheless, we remark that the minimum length modifies the spatial curvature.

\section{Conclusions}

\label{sec5}

In this piece of work, we examined the appearance of a minimum length that
modifies the Heisenberg algebra in the gravitational field equations
describing the Szekeres spacetimes. The existence of the minimum length leads
to the GUP and consequently to new commutation relations. As a result, the
classical limite specifically, Hamilton's equations are modified accordingly.

The Szekeres system is a set of ordinary differential equations that describes
the evolution of the kinematic quantities for a family of inhomogeneous
cosmologies, which are part of the so-called silent universes. The resulting
field equations form an integrable Hamiltonian system. By introducing a
deformed Heisenberg algebra, we were able to define the modified Szekeres
system and study the physical parameters. We found that acceleration can occur
related to the definition of the GUP.

We considered the simplest GUP theory, where the Heisenberg algebra is
modified with a constant scale. This modification leads to a set of
Hamiltonian equations where the integrability properties and the solutions of
the original system are preserved. Nevertheless, when deriving the modified
Szekeres system, we determined the physical parameters and found that they
depend on the deformation parameter.

To understand the effects of GUP on the solution space, we investigated the
phase space of the modified Szekeres system by introducing dimensionless
variables. The asymptotic solutions describe geometries that depend on the
deformed parameter. Specifically, the spatial curvature and the deceleration
parameter are modified, leading to new behaviors that describe cosmic
acceleration. Finally, we discuss the case where the cosmological constant is
introduced into the Szekeres geometry, where we found corrections in the
spatial curvature of the spacetime related to the GUP.

In the future, we plan to extend the present analysis by introducing a
specific deformed algebra and also extend this analysis to the case of the
Extended Uncertainty Principle \cite{eup1}.

\begin{acknowledgments}
The author thanks the support of Vicerrector\'{\i}a de Investigaci\'{o}n y
Desarrollo Tecnol\'{o}gico (Vridt) at Universidad Cat\'{o}lica del Norte
through N\'{u}cleo de Investigaci\'{o}n Geometr\'{\i}a Diferencial y
Aplicaciones, Resoluci\'{o}n Vridt No - 098/2022.
\end{acknowledgments}


\begin{thebibliography}{99}                                                                                               %


\bibitem {Aref1}A.A. Starobinsky, Phys. Lett. B 91, 99 (1980)

\bibitem {guth}A. Guth, Phys. Rev. D 23, 347 (1981)

\bibitem {st1}V. Muller, H.-J. Schmidt and A.A. Starobinsky, Phys. Lett. B
202, 2, 198 (1988)

\bibitem {st2}L.A Kofman, A.D. Linde and A.A. Starobinsky, Phys. Lett. B 157,
5-6, 361 (1985)

\bibitem {w1}R. Wald, Phys.\ Rev. D 28, 2118 (1983)

\bibitem {nh1}G.W. Gibbons and S.W Hawking, Phys. Rev. D 15, 2738 (1977)

\bibitem {nh2}S.W.\ Hawking and J.G. Moss. Phys. Lett. B 110, 35 (1982)

\bibitem {f1}K. Sato, MNRAS 195, 467 (1981)

\bibitem {f2}J.D\ Barrow and A. Ottewill, J. Phys. A\ 16, 2757 (1983)

\bibitem {ba1}J.D. Barrow and J. Stein-Schabes, Phys.\ Lett. A 103, 6-7, 315 (1984)

\bibitem {sim1}K. Bolejko, Phys.\ Rev. \ D 97, 083515 (2018)

\bibitem {mm1}N. Mustapha, G.F.R. Ellis, H. van Elst and M. Marklund, Class.
Quantum\ Grav. 17, 3135 (2000)

\bibitem {ltb2}P.S. Apostolopoulos, N. Brouzakis, N.\ Tetradis and E. Tzavara,
JCAP 0606, (2006)

\bibitem {ltb3}D. Garfinkle, Class. Quantum Grav. 23, 4811 (2006)

\bibitem {novikov}I.D. Novikov, Astron. Zn. 40, 772 (1963)

\bibitem {tm}I. Georg and C.\ Hellaby, Phys. Rev. D 95, 124016 (2017)

\bibitem {mi1}M. Bruni, S. Matarrese and O. Pantano, Astrophys. J. 445, 958 (1995)

\bibitem {bon}W.B. Bonnor, Commun. Math. Phys. 51, 191 (1976)

\bibitem {Gol}A.N. Golubiantnikov and L.M.Truskinovskii, Prikl. Matem. Mekhan.
45, 956 (1981)

\bibitem {grpan1}J.D.Barrow and J. Silk, Astrophys. J. 250, 432 (1981)

\bibitem {grpan2}J.D. Barrow and G. G\"{o}tz, Class. Quantum Gravity 6, 1253 (1989)

\bibitem {zeldpan}Y.B. Zeldovich, Astronomy and Astrophysics. 5, 84 (1970)

\bibitem {szafron}D.A. Szafron, J. Math. Phys. 18, 1673 (1977)

\bibitem {kraj}A. Krasinski, Inhomogeneous Cosmological Models, Cambridge
University Press, Cambridge (2010)

\bibitem {ml1}K. Konishi, G. Paffuti and P. Provero, Phys.\ Lett. B 234, 276 (1990)

\bibitem {ml2}A. Camelia, Int. J. Mod. Phys. D \textbf{11}, 35 (2002).

\bibitem {ml3}P. Martinetti, F. Mercati and L. Tomassini, Rev. Math. Phys. 24,
1250010 (2012)

\bibitem {ml4}Ashtekar, A. and Lewandowski, J., Class. Quantum Grav., 21, R53, (2004)

\bibitem {Maggiore}M. Maggiore, Phys. Lett. B 304, 65 (1993)

\bibitem {sb1}S. Hossenfelder, Living Reviews in Relativity 16, 2, (2013)

\bibitem {sb2}S.F. Hassan and M.S. Sloth, Nucl. Phys. B 674, 434 (2003

\bibitem {sb3}L. Sriramkumar and S. Shankaranarayanan, JHEP 0612, 050 (2006)

\bibitem {sb4}G.A. Palma and S.P. Patil, JHEP 0904, 005 (2009)

\bibitem {gg1}A. Das, S. Das, N.R. Mansour and E.C. Vagenas, Phys. Lett. B
819, 136429 (2021)

\bibitem {gg2}L.A. Escamilla, D. Fiorucci, G. Montani and E. Di Valentino

\bibitem {gg3}S.-T. Hong, Y.-W. Kim and Y.-J. Park, Mod. Phys. Lett. A 37,
2250186 (2022)

\bibitem {gg4}A.W. Khanday, S. Upadhyay and P.A. Ganai, Eur. Phys. J. C 82,
1164 (2022)

\bibitem {gg5}A. Sahoo, S.K. Tripathy and B. Mishra, Eur. Phys. J. \ C 84, 325\ (2024)

\bibitem {gg6}H. Hoshimov, O. Yunusov, F. Atamurotov, M. Jamil and A.
Abdujabbarov, Phys. Dark Univ. 43, 101392 (2024)

\bibitem {gg7}S. Segreto and G. Montani, Eur. Phys. J. C 84, 796 (2024)

\bibitem {gg8}A. Giacomini, G. Leon, A. Paliathanasis and S.\ Pan, Eur. Pjhys.
J. C 80, 931 (2020)

\bibitem {gg9}A. Paliathanasis, G. Leon, W. Khyllep, J. Dutta and S. Pan, Eur.
Phys. J. C 81, 607 (2021)

\bibitem {gg10}A. Paliathanasis, S. Pan and S. Pramanik, Class. Quantum Grav.
32, 245006 (2015)

\bibitem {gg11}L. Buoninfante, G. Lambiase, G.G. Luciano and L. Petruzziello,
Eur. Phys.\ J. C 80, 853 (2020)

\bibitem {anszek}A. Paliathanasis and P.G.L. Leach, Phys. Lett. A\ 381, 1277 (2017)

\bibitem {szek0}P. Szekeres, Commun. Math. Phys. 41, 55 (1975)

\bibitem {Vagenas}S. Das and E.C. Vagenas, Phys. Rev. Lett. 101, 221301 (2008)

\bibitem {Moayedi}S.K. Moayedi, M.R. Setare and H Moayeri, Int. J. Theor.
Phys. 49, 2080 (2010)

\bibitem {hm1}E. Vagenas, A.F. Ali, M. Hemeda and H. Alshal, Eur. J. Phys. C
79, 398 (2019)

\bibitem {hm2}P. Bosso, Phys.\ Rev.\ D 97, 126010 (2018)

\bibitem {hm3}R. Casadio and F. Scardigli, Phys. Lett. B 807, 135558 (2020)

\bibitem {hm4}V. Nenmeli, S. Shankaranarayanan, V. Todorinov and S. Das, Phys.
Lett. B 821, 136621 (2021)

\bibitem {hm5}F. Scardigli, M. Blasone, G. Luciano and R. Casadio, Eur. J.
Phys. C 78, 728 (2018)

\bibitem {hm6}S. Segreto and G. Montani, Eur. J. Phys. C 83, 385 (2023)

\bibitem {hm7}K. Blanchette, S. Das and S. Ratgoo, J. High. Energ. Phys. 2021,
62 (2021)

\bibitem {hm8}M. Maggiore, Phys. Lett. B 319, 83 (1993)

\bibitem {eup1}A. Kempf, G. Mangano and R. B. Mann, Phys. Rev. D 52, 1108 (1995)
\end{thebibliography}
\end{document}